\documentclass[aps,epsfig]{revtex4} 

\usepackage[dvips]{graphicx} 
\usepackage{amssymb}
\usepackage{subfigure}

\newcommand{\be}{\begin{equation}}
\newcommand{\ee}{\end{equation}}
\newcommand{\bea}{\begin{eqnarray}}
\newcommand{\eea}{\end{eqnarray}}
\newcommand{\fnl}{f_{\rm NL}}
\newcommand{\alm}{a_{\ell m}^X}
\newcommand{\almL}{a_{\ell m}^{X,\rm L}}
\newcommand{\almNL}{a_{\ell m}^{X,\rm NL}}

\begin{document}

\title{Temperature and polarization CMB maps from primordial
non-Gaussianities of the local type}

\author{Michele Liguori$^1$,  Amit Yadav$^2$, Frode K. Hansen$^3$, Eiichiro Komatsu$^4$, Sabino
  Matarrese$^5$, Benjamin Wandelt$^2$}

\affiliation{$^1$Department of Applied Mathematics and Theoretical Physics, Centre for
Mathematical Sciences, University of Cambridge, Wilberfoce Road, Cambridge, CB3 0WA, United
Kingdom}

\affiliation{$^2$Department of Astronomy,  University of Illinois at Urbana-Champaign, 
1002 W.~Green Street, Urbana, IL 61801}

\affiliation{$^3$Institute of Theoretical Astrophysics, University of Oslo, 
P.O. Box 1029 Blindern, 0315 Oslo, Norway} 

C\affiliation{$^4$Department of Astronomy, University of Texas at
Austin, 2511 Speedway, RLM 15.30 6, TX 78712}

\affiliation{$^5$Dipartimento di Fisica ``G. Galilei'', Universit\`a di Padova and 
INFN, Sezione di Padova, via Marzolo 8, I-35131, Padova, Italy}

\date{\today}

\begin{abstract} 
  The forthcoming {\em Planck} experiment will provide 
  high sensitivity polarization measurements that will allow us to
  further tighten the $\fnl$ bounds from the temperature data. 
  Monte Carlo simulations of non-Gaussian CMB maps have been used as 
 a fundamental tool to characterize non-Gaussian signatures in the data,
 as they allow us 
  to calibrate any statistical estimators and understand the effect
  of systematics, foregrounds and other contaminants.   
  We describe an algorithm to generate high-angular resolution
  simulations of non-Gaussian CMB maps in temperature {\it and
  polarization}. We consider non-Gaussianities of the local type, for
  which the level of non-Gaussianity is defined by the dimensionless
  parameter, $\fnl$. 
  We then apply the temperature and polarization fast cubic statistics 
  recently developed by Yadav {\em et al.} to a set of non-Gaussian 
  temperature and polarization simulations. 
  We compare our results to theoretical expectations 
  based on a Fisher matrix analysis, test the unbiasedness 
  of the estimator, and study the dependence of the error bars on
  $\fnl$. All our results are in very good agreement with theoretical 
  predictions, thus confirming the reliability of both the simulation
  algorithm and the fast cubic temperature and polarization estimator.
\end{abstract}

\maketitle

\section{Introduction}

Small, but non-vanishing 
non-Gaussianity of primordial cosmological perturbations is a general
prediction of inflation. 
The amplitude of 
the expected non-Gaussian signal is model-dependent and can vary by
many orders of magnitude from one inflationary scenario to another. For example, the
non-Gaussian signatures produced by single-field slow-roll inflation models
are tiny and far below the present and forthcoming experimental
sensitivity \cite{acqua,maldacena}. On the other hand many other scenarios predict a level of
non-Gaussianity that is within reach of present
and forthcoming experiments like WMAP and {\em Planck} (see
e.g. \cite{Lyth,BMR,komrev,Ginfl,DBI1,DBI2,Shellard,multifield}). For this
reason an experimental detection of non-Gaussianity would rule out 
the simplest scenarios of slow-roll inflation. More in general,
experimental bounds on primordial non-Gaussianity allow us to
significantly constrain different scenarios 
for the generation of perturbations in the context of primordial inflation.

Primordial non-Gaussianity from inflation can be described in terms of
the 3-point correlation function of the curvature perturbations,
$\Phi(\mathbf{k})$, in Fourier space: 

\be
\langle \Phi(\mathbf{k_1}) \Phi(\mathbf{k_2}) \Phi(\mathbf{k_3})
\rangle = (2\pi)^3 \delta^{(3)}(\mathbf{k_1 + k_2 +
  k_3})F(k_1,k_2,k_3) \; .
\ee

Note that $\Phi$ is the curvature perturbation during the matter era,
and temperature anisotropy in the Sachs-Wolfe limit is given by 
$\Delta T/T=-\Phi/3$.
Depending on the shape of the function $F(k_1,k_2,k_3)$, we can divide
non-Gaussianity from inflation into two different classes: 
local non-Gaussianity, where $F$ is large for
{\em squeezed} configurations (i.e. configurations in which $k_1 <<
k_2, k_3$), and non-local non-Gaussianity of the {\em equilateral} type, where
the largest contributions come from modes with $k_1 \sim k_2 \sim
k_3$. The former kind of non-Gaussianity can be produced in models
where primordial perturbations are not generated by inflaton
itself but by a second light scalar field (like e.g. in the curvaton
model).
The latter comes from single field models with a non-minimal
Lagrangian containing higher derivative operators. 
In this paper we will focus on non-Gaussianity of the local type,
where the primordial curvature perturbation $\Phi$ 
can be described in terms of the following real space parameterization:

\be\label{eqn:phiNG}
\Phi(\mathbf{x}) = \Phi_L(\mathbf{x}) + \fnl \left( \Phi_L^2(\mathbf{x}) - \langle \Phi_L^2(\mathbf{x}) \rangle
\right) \;.
\ee

In the last formula $f_{\rm NL}$ is a parameter that defines the amplitude
of the primordial non-Gaussian signal. Our previous statement about
the detectability of non-Gaussian signatures from inflation can be
precisely quantified in terms of this parameter. In standard scenarios of
single-field slow roll inflation $f_{\rm NL}$ is generally predicted to be
very small and undetectable ($\sim 10^{-2}$ at the end of inflation, $\sim 1$ when 
second order perturbation theory after inflation is taken into
account) whereas other scenarios, like the curvaton or variable
decay width models, can naturally give rise to relatively large values
of $\fnl$ ($\fnl \sim 10$). This justifies the claim
that an experimental 
detection of $\fnl$ would rule out the simplest single-field 
inflationary paradigm and allow us to put significant constraints on
the other inflationary scenarios. 

The best way to put experimental 
bounds on $\fnl$ is
to look for non-Gaussianities in CMB anisotropies (but it has been
recently pointed out that future deep galaxy-surveys and 21 cm
background measurements could provide promising results 
\cite{Sefusatti,Pillepich,Cooray}). 
The most stringent constraints on $\fnl$
so far come from measurements of the CMB angular bispectrum on the
WMAP temperature data 
$
-36 < \fnl < 100 \;\;  (95 \% \, c.l.) 
$
\cite{NGWMAP1,NGWMAP3,Creminelli2}.
This constraint corresponds to a $1-\sigma$ error of $\Delta \fnl =
34$.  
A Fisher matrix analysis by the authors of \cite{KS2001} showed that
WMAP will in principle be able to reach $\Delta \fnl = 20$, while the 
forthcoming {\em Planck} satellite can achieve $\Delta \fnl = 5$. This means 
that {\em Planck} will be sensitive to the level of non-Gaussianity
predicted by a vast range of different inflationary models. We can
improve this constraint further by including
the polarization data.
For WMAP all the non-Gaussian information is basically 
contained in the temperature data, due to large errors in 
polarization measurements. {\em Planck}, on the other hand, 
will characterize polarization
fluctuations with high accuracy. This will allow us to exploit the additional
information contained in polarization data and to gain a further
factor of order $2$ in $\Delta \fnl$, thus yielding $\Delta \fnl
\simeq 3$ \cite{BabichZalda}.
A crucial step in order to exploit all the information contained in
the future {\em Planck} dataset is then to {\em extend the tools previously 
developed for temperature non-Gaussianity in order to include polarization}. 
This program has been recently started by the authors of \cite{Yadav},
where the fast cubic statistic used to analyze WMAP temperature
data \cite{KSW,CreminelliKSW} was taken as a starting point to build an 
optimal cubic estimator that is sensitive to
a combination of temperature and polarization primordial
fluctuations. In this paper we will extend the non-Gaussian analysis 
toolkit in order to include the second fundamental element: Monte Carlo
simulations of primordial non-Gaussian polarized CMB maps.

In section \ref{sec:maps} we will summarize the
original algorithm and describe its extension to polarization. We will 
then apply the fast cubic statistic of \cite{Yadav} to a set of
polarized non-Gaussian maps.
In this paper we will manly focus our attention on map
generation, so the purpose for applying the estimator is mainly to
check the reliability of the final maps. This will be done by
comparing the final outputs
to theoretical predictions in ideal conditions. However in a
forthcoming publication we will describe how we actually used the maps
in order to test, calibrate and optimize the estimator.

\begin{figure}[h]
\begin{center}
\includegraphics[height=0.6\textheight,width = 0.8\textwidth]{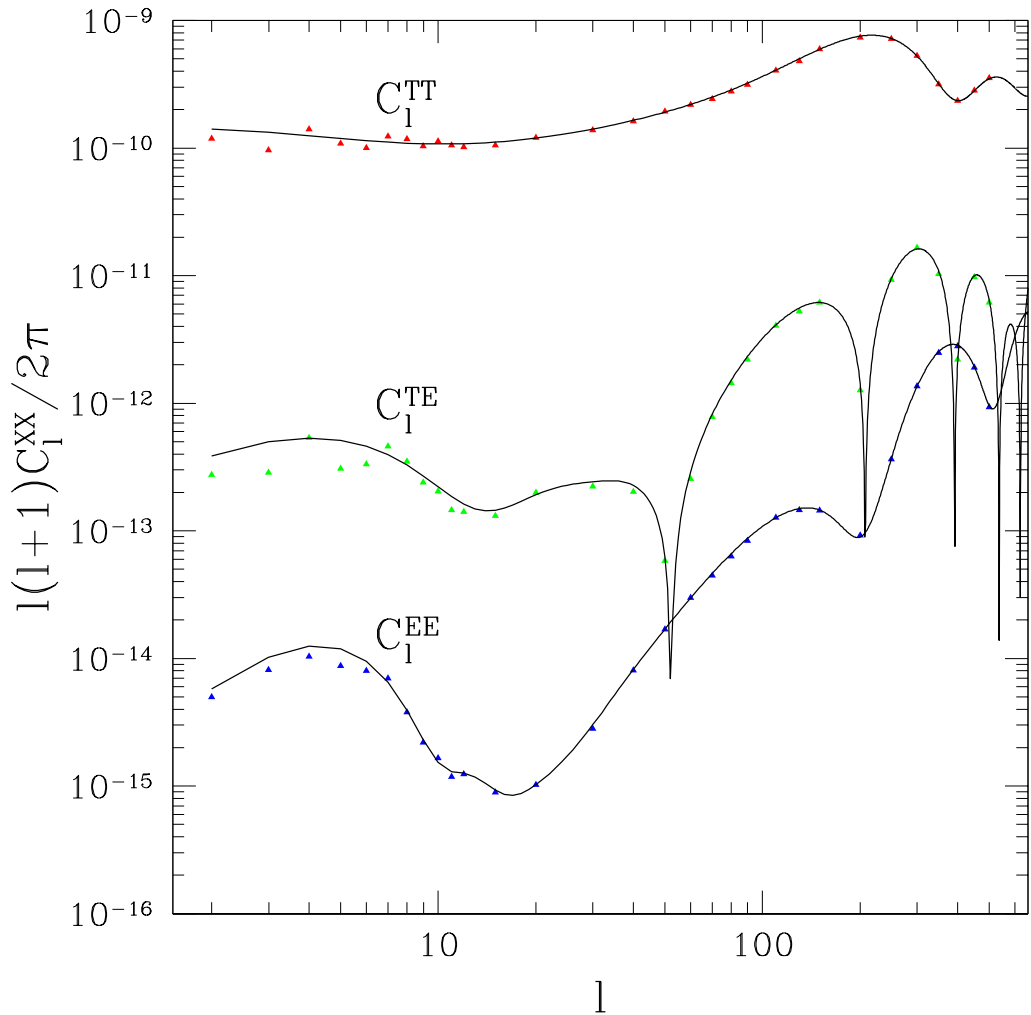}
\caption{CMB angular power spectra extracted from $10$ simulations
(triangles)
are compared to the theoretical ones computed with
CMBfast for the same model (solid black lines) . 
The cosmological parameters are
$\Omega_b = 0.042$, $\Omega_{cdm} = 0.239$, $\Omega_L = 0.719$, $h = 0.73$
$n = 1$, and $\tau = 0.09$ (same for all the following figures, unless
otherwise stated).
}\label{fig:cl}
\end{center}
\end{figure}

\begin{figure}[h]
\begin{center}
\includegraphics[height = 0.5\textheight, width = 0.7\textwidth]{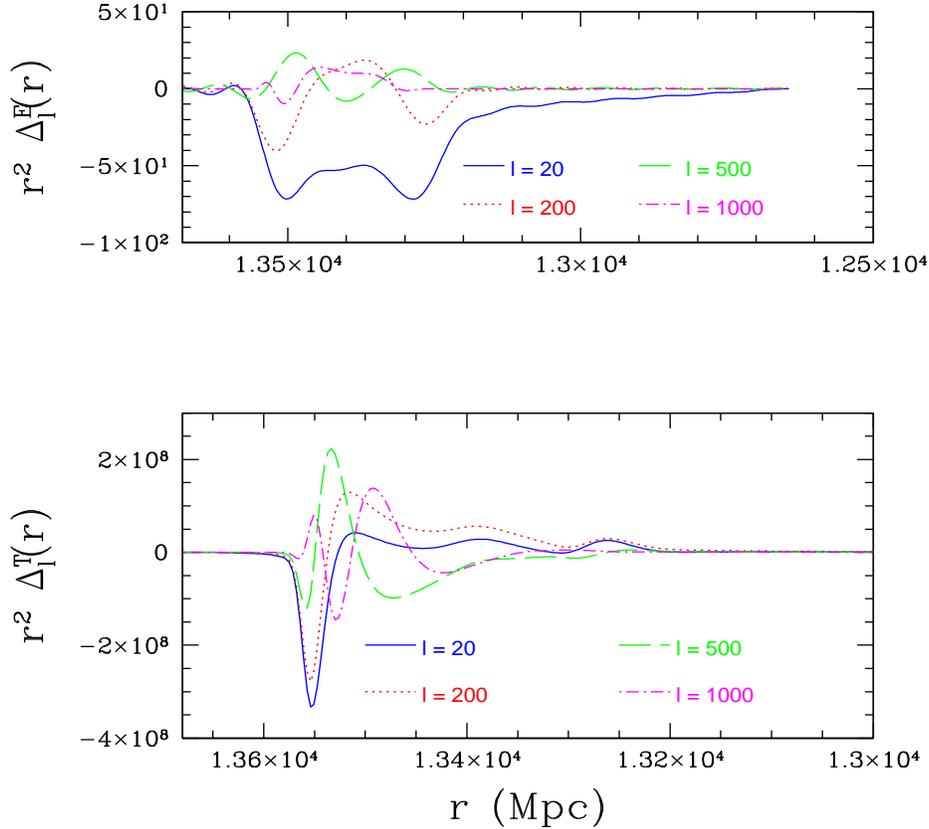}
\caption{Temperature (bottom panel) and polarization (upper panel)
  transfer functions at high $\ell$ at last scattering.}\label{fig:rtf1}
\end{center}
\end{figure}

\begin{figure}[t]
\begin{center}
\includegraphics[height = 0.5\textheight, width = 0.7\textwidth]{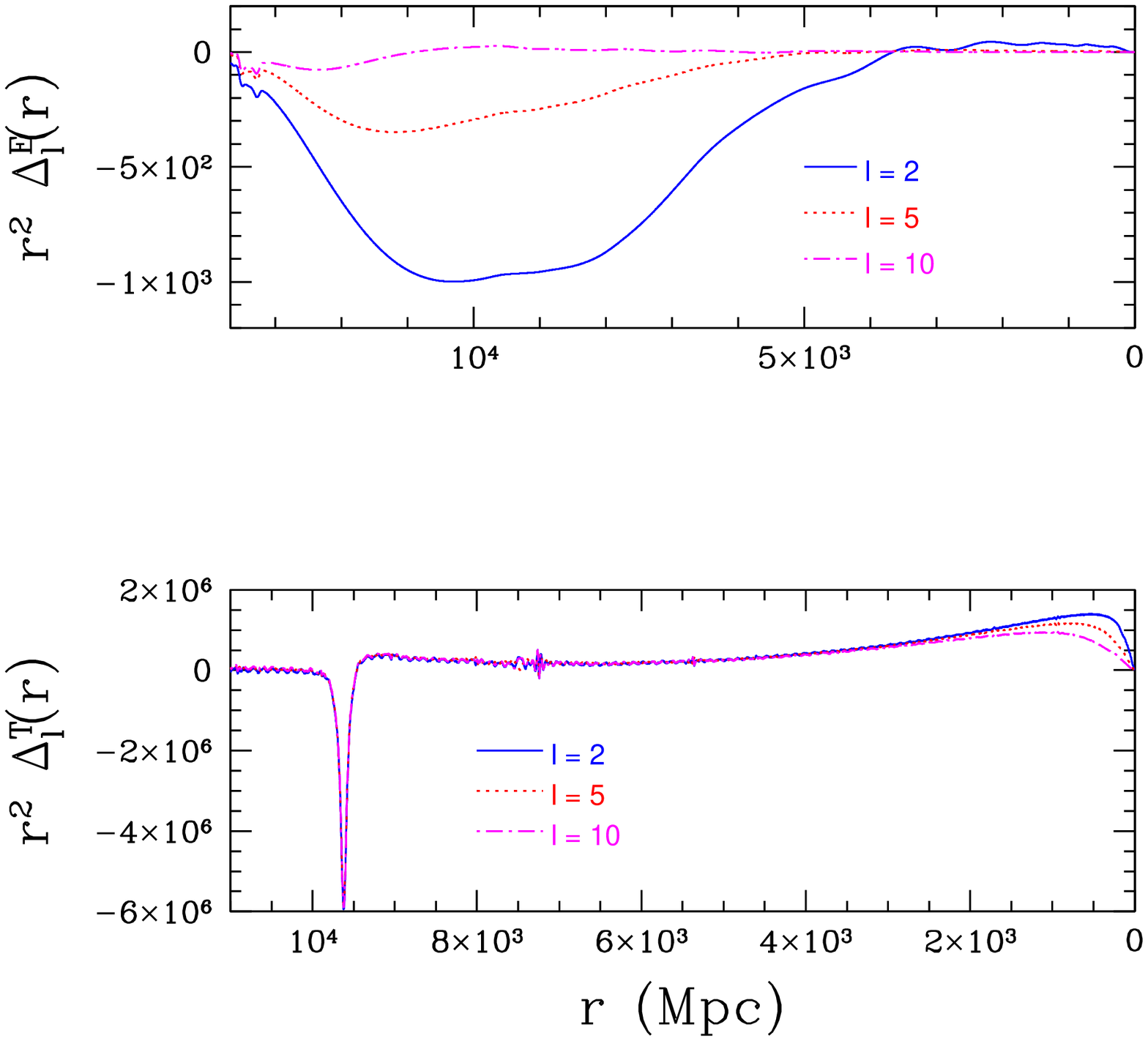}
\caption{Temperature (bottom panel) and polarization (upper panel)
  transfer functions at low $\ell$ (reionization and late ISW
  contributions are visible). The oscillations visible in the plots
  are little numerical artifacts which have negligible impact on the final
  results. We have explicitly checked this by increasing the
  resolution in the k and r-grid by factors of $2$ and $4$ without
  noticing any improvement in the accuracy of the final $C_\ell$, that
  can be already reconstructed well using the sampling chosen in the
  paper (see fig. \ref{fig:cl}) }\label{fig:rtf3}
\end{center}
\end{figure}

\section{Generation of polarized non-Gaussian CMB maps}\label{sec:maps}

\begin{figure}[t]
\begin{center}
\includegraphics[height = 0.5\textheight, width = 0.7\textwidth]{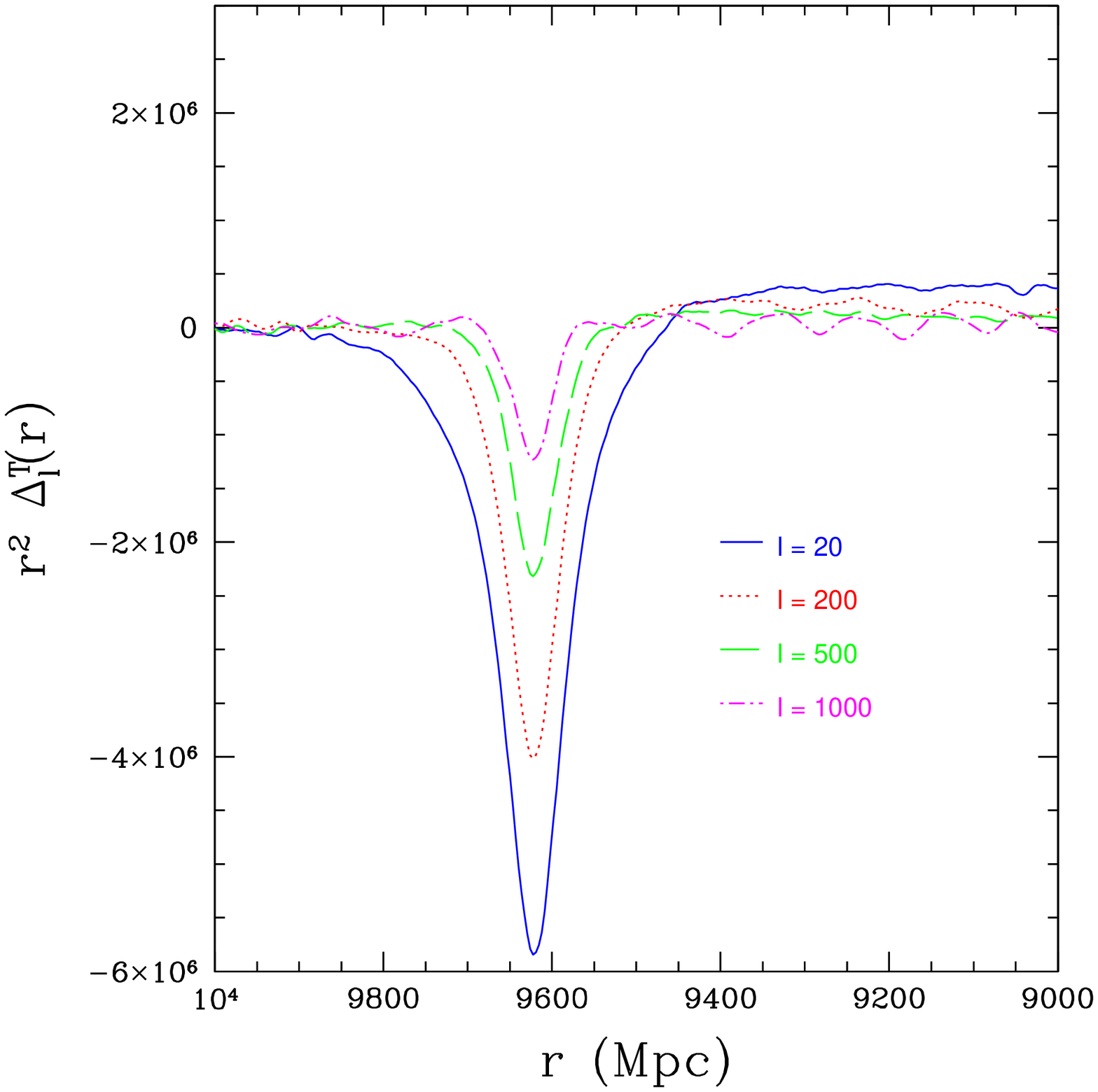}
\caption{Temperature
  transfer functions at high $\ell$ and $r$ corresponding to the epoch of
  reionization. Polarization transfer functions at large $\ell$ are
  zero in this range.}\label{fig:rtf2}
\end{center}
\end{figure}

Realistic
simulations of non-Gaussian CMB maps are indispensable tools
 for measurements
of non-Gaussian signals in the data, as they allow us to test
and calibrate estimators and also to include and study all the spurious
non-Gaussian signals introduced by contaminants like foregrounds,
secondary anisotropies, instrumental noise and so on.   
  
The first simulations of temperature maps with primordial 
non-Gaussianity from $\fnl$ were carried out by Komatsu et
al., and used extensively to study Gaussianity of 
the WMAP data  \cite{NGWMAP1} 
as well as non-trivial topology of the universe \cite{Cornish}.
Then, Liguori et al.\cite{liguori} have succeeded in increasing the
computational speed, reducing the memory requirement and, most
importantly, improving accuracy of the simulated temperature maps.
We take this new algorithm developed in \cite{liguori} as a starting point. 

Our starting point is the relation between the
primordial curvature perturbation $\Phi$ and the CMB multipoles
$\alm$ via radiation transfer functions $\Delta^{\rm X}_\ell$. 

\be
\alm = \int \frac{d^3 k}{(2 \pi)^3} \Phi(\mathbf{k}) Y_{\ell
  m}(\hat{k}) \Delta_\ell^X(k) \; ,
\ee
where $\rm X$ refers to either the temperature component $\rm T$ 
or the polarization component $\rm E$.

The kind of non-Gaussianity we are considering has a very simple form
in real space, where it is local and
the non-Gaussian part of the curvature perturbation is simply the square of the
Gaussian part (see formula \ref{eqn:phiNG}). For this reason it is
convenient to work in real space and define the real space transfer
functions $\Delta_\ell(r)$ as:

\be\label{eqn:rtf}
\Delta_\ell^X(r) \equiv \frac{2}{\pi} \int dk k^2 j_\ell(kr) \Delta_\ell^X(k) \; ,
\ee 
where $j_\ell(kr)$ is the spherical Bessel function of order
$\ell$. It can be shown that $\Delta_{\ell}(r)$
links the primordial curvature perturbation $\Phi(\mathbf{r})$ in real
space to the $\alm$ through the following relation \cite{NGWMAP1,liguori}

\be\label{eqn:phi2alm}
\alm = \int dr r^2 \Delta_\ell^X(r) \Phi_{\ell m}(r) \; .
\ee

In this last formula we have introduced the quantities $\Phi_{\ell
  m}(r)$, which represent the spherical harmonic expansion multipoles  
of the curvature perturbation $\Phi(r,\hat{r})$ on a shell of given
  radius $r$. In formulae: 

\be\label{eqn:philmdef}
\Phi_{\ell m}(r) = \int d\Omega_{\hat{r}} Y_{\ell m}(\hat{r}) \Phi(r,\hat{r}) \; .
\ee

We define the radius $r$ as $r = c(\tau_0 -\tau)$, where $c$ is the
speed of light and $\tau_0 - \tau$ is the lookback conformal time. The
radius $r$ varies from the origin $r = 0$ to the present time cosmic
horizon $r = c\tau_0$. The radii in which $\Phi_{\ell m}(r)$ must be
generated depend on the features of the real space transfer function
$\Delta_{\ell}^X(r)$ in equation (\ref{eqn:phi2alm}). We will come back
to this shortly.

Let us assume for the moment that we have been able to numerically
generate the Gaussian part of the curvature perturbation multipoles
$\Phi^L_{\ell m}(r)$ 
for the chosen set of radii. Starting from here we can
now generate the non-Gaussian
part $\Phi^{NL}_{\ell m}(r)$ in the following way. First of all we
  harmonic transform $\Phi^L_{\ell m}(r)$ to get the gaussian part of the 
curvature perturbation in real space:

\be
\Phi_{\rm L}(r,\hat{r}) = \sum_\ell \sum_m \Phi^L_{\ell m}(r) Y_{\ell
  m}(\hat{r}) \; .
\ee

Then we square $\Phi_{\rm L}(r,\hat{r})$ to get the non Gaussian part of the 
curvature perturbation on each sampled spherical shell: 
$\Phi_{\rm NL}(r,\hat{r}) \equiv \Phi_{\rm L}^2(r,\hat{r}) - \langle
\Phi_{\rm L}^2(r,\hat{r}) \rangle$. We then calculate the multipoles of this 
non-Gaussian part through a backward harmonic transform:

\be
\Phi^{\rm NL}_{\ell m}(r) \equiv \int d\Omega_{\hat{r}}
\Phi_{\rm NL}(r,\hat{r}) Y_{\ell m}(\hat{r}) \; .
\ee

Having computed $\Phi^L_{\ell m}(r)$ and $\Phi^{NL}_{\ell m}(r)$ we 
can finally obtain the Gaussian and non-Gaussian part of the CMB
multipoles, $\almL$ and $\almNL$ respectively, by applying formula 
(\ref{eqn:phi2alm}):

\bea
\almL & = & \int dr r^2 \Delta_\ell^X(r) \Phi^{\rm L}_{\ell m}(r) \\
\almNL & = &  \int dr r^2 \Delta_\ell^X(r) \Phi^{\rm NL}_{\ell m}(r) \; .
\eea

A CMB map for a chosen value of $\fnl$ can then be obtained simply 
by summing $\almL + \fnl \almNL$. This means that with a single
generation of $\alm$ and $\almNL$ it is possible to generate maps
for any value of $\fnl$.

We are still left with one problem unsolved i.e. how do we
generate the Gaussian curvature perturbation multipoles $\Phi^L_{\ell
m}(r)$ ?
This issue is complicated by the fact that curvature perturbation multipoles are 
correlated in real space. 
The obvious solution would be to generate curvature perturbations in 
Fourier space,  $\Phi({\mathbf k})$, Fourier transform back to real space
to obtain $\Phi({\mathbf x})$, change the coordinates from Cartesian
to polar to obtain $\Phi(r,\hat{n})$, and finally harmonic transform
to obtain $\Phi_{\ell m}(r)$. 
This is the original approach taken by \cite{NGWMAP1}, which
is computationally quite expensive. Also, the coordinate transformation
from Cartesian to polar limits accuracy of the maps, especially at 
high multipoles.

A novel approach developed in \cite{liguori} solves this issue by 
generating $\Phi_{\ell m}(r)$ {\it directly}, without ever worrying
about the coordinate transformation.
It has been shown in \cite{liguori} that the $\Phi_{\ell m}(k)$ and
$\Phi_{\ell m}(r)$ are related by a spherical Bessel transform:

\be\label{eqn:bt}
\Phi_{\ell m}(r) = \frac{(-i)^\ell}{2 \pi} \int dk k^2 j_{\ell}(kr)
  \Phi_{\ell m}(k) \; .
\ee

The problem with this expression is that the Bessel functions
oscillate very rapidly. This implies that, for each $(\ell, r)$, 
the integral above must be
sampled in many different $k$ in order to attain sufficient accuracy,
thus making the computational cost of such an algorithm prohibitive. A
much more convenient solution was found in \cite{liguori}; the idea is
to start with a set of Gaussian independent ``white noise'' coefficients
$n_{\ell m}(r)$ characterized by the following correlation function:

\begin{equation}
\label{eqn:whitenoise} 
\left \langle n_{\ell_1 m_1}(r_1)
n^*_{\ell_2 m_2}(r_2) \right \rangle = 
\frac{\delta^D(r_1-r_2)}{r^2}\delta_{\ell_1}^{\ell_2} \delta_{m_1}^{m_2}\; ;  
\end{equation} 

it can be now shown that Gaussian curvature perturbation multipoles $\Phi_{\ell
  m}^{\rm L}(r)$ with the right correlation properties can be
obtained through a convolution of the $n_{\ell m}$ coefficients with
suitable ``filters'' $W_\ell$:

\begin{equation}
\label{eqn:nlm2phil} 
\Phi^{\rm L}_{\ell m}(r) = \int \! dr_1 \, r_1^2 \, n_{\ell m}(r_1) 
W_\ell(r,r_1) \; ,
\end{equation} 
where the functions $W_\ell$ are defined as

\begin{equation}
\label{eqn:filter} 
W_\ell(r,r_1) =
\frac{2}{\pi} \int \! dk \, k^2 \, \sqrt{P_\Phi(k)} \, j_\ell(kr)
j_\ell(kr_1) \; ,  
\end{equation}
and $P_\Phi(k)$ is the power spectrum of the primordial curvature perturbation 
$\Phi_{\rm L}(\mathbf{k})$.
As depicted in Fig. \ref{fig:wl1}, \ref{fig:wl2} the filter functions $W_\ell$
are smooth. Moreover, as also suggested by the Limber approximation applied
to equation (\ref{eqn:filter}), $W_\ell(r,r_1)$ is narrowly peaked around $r$ when $l
\gtrsim 10$. This allows to sample the integral
(\ref{eqn:nlm2phil}) in much less points than it would be
required for the Bessel transform (\ref{eqn:bt}), thus making the
problem computationally feasible. Obviously the problem of sampling a
highly oscillatory integrand has not disappeared completely, but it has been
reduced to the generation of $W_\ell(r,r_1)$. A trick here
is that the filters $W_\ell(r,r_1)$ can be {\em pre-computed and stored}
once and for all for a given cosmological model and their calculation does not enter in
the actual Monte Carlo simulation algorithm. The same argument applies
to the radiation transfer functions $\Delta_\ell(r)$ defined in (\ref{eqn:rtf}).

Non-Gaussian temperature maps
produced with the algorithms described in this section had been
already described in \cite{liguori}. Adding polarization to those maps
is conceptually straightforward: all one needs to do 
is to replace $X = T$ with $X = E$ in the previous expressions. This
amounts to generating the primordial curvature perturbation in exactly 
the same way for temperature and
polarization maps and finally to use polarization transfer functions
in place of temperature transfer functions in the line of sight
integral (\ref{eqn:phi2alm}) in order to get $a_{\ell m}^E$. Despite
its conceptual immediateness, including polarization in the maps
is not technically straightforward. The reason is that CMB
polarization is produced by different physical mechanisms with respect 
to those producing CMB temperature anisotropies. 
The polarization transfer functions $\Delta^E_\ell(r)$ 
present then several differences with respect to $\Delta^T_\ell(r)$
and must be sampled in a different way, thus changing sampling regions
and discretization of the r-coordinate which appears in
$\Phi_{\ell m}(r)$, $\Delta_\ell(r)$, $W_\ell(r,r_1)$. These
technical details will be illustrated in the following two sections.

\subsection{Real space transfer functions}\label{sec:rstf}

The cosmological model we chose to generate our non-Gaussian maps is
characterized by the following parameters: $\Omega_{cdm} = 0.239$, 
$\Omega_b = 0.042$, $\Omega_\Lambda =
0.719$, $\tau = 0.09$, $h = 0.73$. We considered both a scale
invariant primordial spectral index $n=1$ and $n=0.95$, the latest one
being the WMAP 3-years best-fit value \cite{NGWMAP3}. Starting from these
parameters we generate and extract the Fourier space radiation 
transfer functions $\Delta^X_\ell(k)$ from a Boltzmann integrator, like for example
CMBfast, and then make the integral (\ref{eqn:rtf}) to get
$\Delta^X_\ell(r)$. The behavior of $\Delta^X_\ell(r)$ reflects the
underlying temperature and polarization CMB physics. In
Fig. \ref{fig:rtf1}, we plot the real space
temperature and polarization transfer functions for several different
values of $\ell > 20$. For the model under examination the conformal time at
last scattering, defined
as the peak of the visibility function, is $\tau_* \simeq 277 \; \rm Mpc$
($c= 1$) while the present cosmic horizon is $\tau_0 \simeq 13682 \; \rm
Mpc$. We thus expect most of the signal to be generated at $r_* \equiv
\tau_0 -\tau_* \simeq
13400 \; \rm Mpc$, consistently with what shown in the figure. Despite
being smaller, contributions at lower redshifts cannot be neglected. We know
that both reionization and the late integrated Sachs Wolfe effect
produce significant contributions, especially at low $\ell$'s. The
reionization signal is particularly important for polarization, as it
produces the observed bump at low $\ell$'s in the polarization
spectrum. This is reflected in the behavior of the temperature and
polarization transfer functions at low $\ell$ in the
post-recombination region, accordingly to what depicted in
Fig. \ref{fig:rtf2} and Fig. \ref{fig:rtf3}. 
According to the radiative transfer
physics contained in $\Delta_\ell(r)$, the last scattering surface
must be sampled using a large
number of points in order to accurately reproduce the acoustic
oscillations in the CMB spectrum, while in the low redshift region a
good accuracy can be reached with a coarser sampling. More details
about the sampled regions and intervals are in table
\ref{tab:rsampling}; the idea was to refine the r-grid until a good
accuracy in the final $C_\ell$ from the simulated map was reached (see
Fig. \ref{fig:cl}). However further sampling optimization in order to
improve the speed of the algorithm is probably possible; an algorithm
aimed at this kind of optimization is described in 
\cite{Smith} in the context of bispectrum estimation.

\bigskip

\begin{table}
\begin{tabular}{|c|c|c|c|}
\hline
{\bf Region} & {\bf Bounds} & {$\mathbf{\Delta r}$} & {\bf N. of shells} \\
\hline
Recombination & $12632 \; {\rm Mpc} < r < 13682 \; {\rm Mpc}$ & $3.5
\rm\; Mpc$ & $300$ \\
\hline
Reionization 1 &  $10007 \; {\rm Mpc} < r < 12632 \; {\rm Mpc}$ & $105
\rm Mpc$ & $25$ \\
\hline
Reionization 2 & $9377 \; {\rm Mpc} < r < 10007 \; {\rm Mpc}$ & $35
\rm Mpc$ & $18$ \\
\hline
Low redshifts & $0 \; {\rm Mpc} < r < 9377 \; {\rm Mpc}$ & $105 \rm
Mpc$ & $89$ \\
\hline
\end{tabular}\caption{Sampling of the
  r-coordinate in different regions of the simulation box. Different 
intervals must be sampled with different resolutions, according to
the radiative transfer physics described in section \ref{sec:rstf}.}
\label{tab:rsampling}
\end{table}

\subsection{Filter functions}

After generating the real radiation transfer functions and fixing the
radial coordinate grid, the $W_\ell(r,r_1)$ functions defined in 
(\ref{eqn:filter}) must be generated for 
each value of $\ell$, $r$. Due to the highly oscillatory nature of
the Bessel functions appearing in the definition of $W_\ell(r,r_1)$,
a large number of points is required when sampling the integrand. 
This makes the numerical computation 
of $W_\ell(r,r_1)$ quite slow. However, as we were already stressing
above, this is not a problem as the $W_\ell(r,r_1)$ functions are
pre-computed and stored before the actual Monte Carlo map generation.
When computing $W_\ell$ it is useful to make the simple substitution 
$t = kr$ in the integrand of (\ref{eqn:filter}). This substitution yields:

\be
W_\ell(r,r_1) = 2 \pi r^{\frac{-n-2}{2}} I_\ell \left( \frac{r_1}{r} \right) \; ,
\ee 

where we have defined:

\be
I_\ell(x) \equiv \int dt t^{\frac{n}{2}} j_l(t) j_l(tx) \; .
\ee 

From the last formulae we see that $W_\ell(r,r_1)$ actually depends
only on the ratio ${r_1/r}$ and not on $r_1$ and $r$ separately. This
allows to reduce the dimensionality of the problem and thus to speed up
the calculations.

\begin{figure}[t!]
\begin{center}
\includegraphics[height=0.5\textheight, width = 0.7\textwidth]{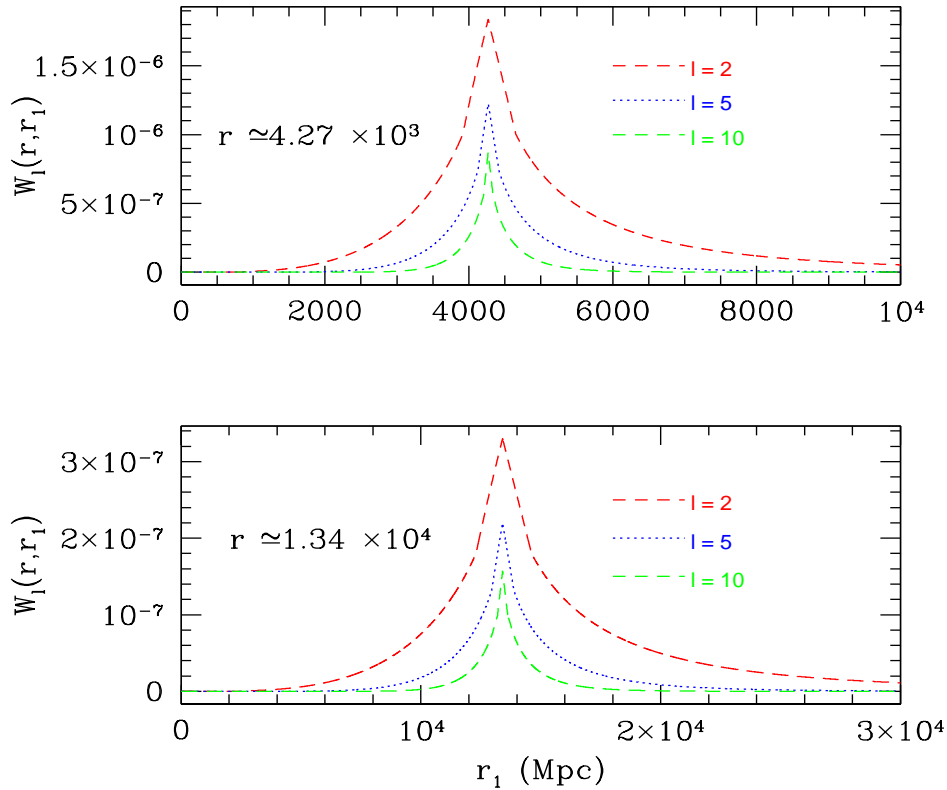}
\caption{Filter functions $W_\ell(r,r_1)$ plotted as a function 
of $r_1$ for two different fixed
  values of $r$. Here we consider low l-values $l \leq 10$, for which
  the $W_\ell(r,r_1)$ are different from zero and must therefore be
  sampled in a large $r_1$ region. At high $\ell$ these functions
  become more and more peaked around $r$, as shown in Fig. \ref{fig:wl2}.}\label{fig:wl1}
\end{center}
\end{figure}

\begin{figure}[t!]
\begin{center}
\includegraphics[height=0.5\textheight,width = 0.7\textwidth]{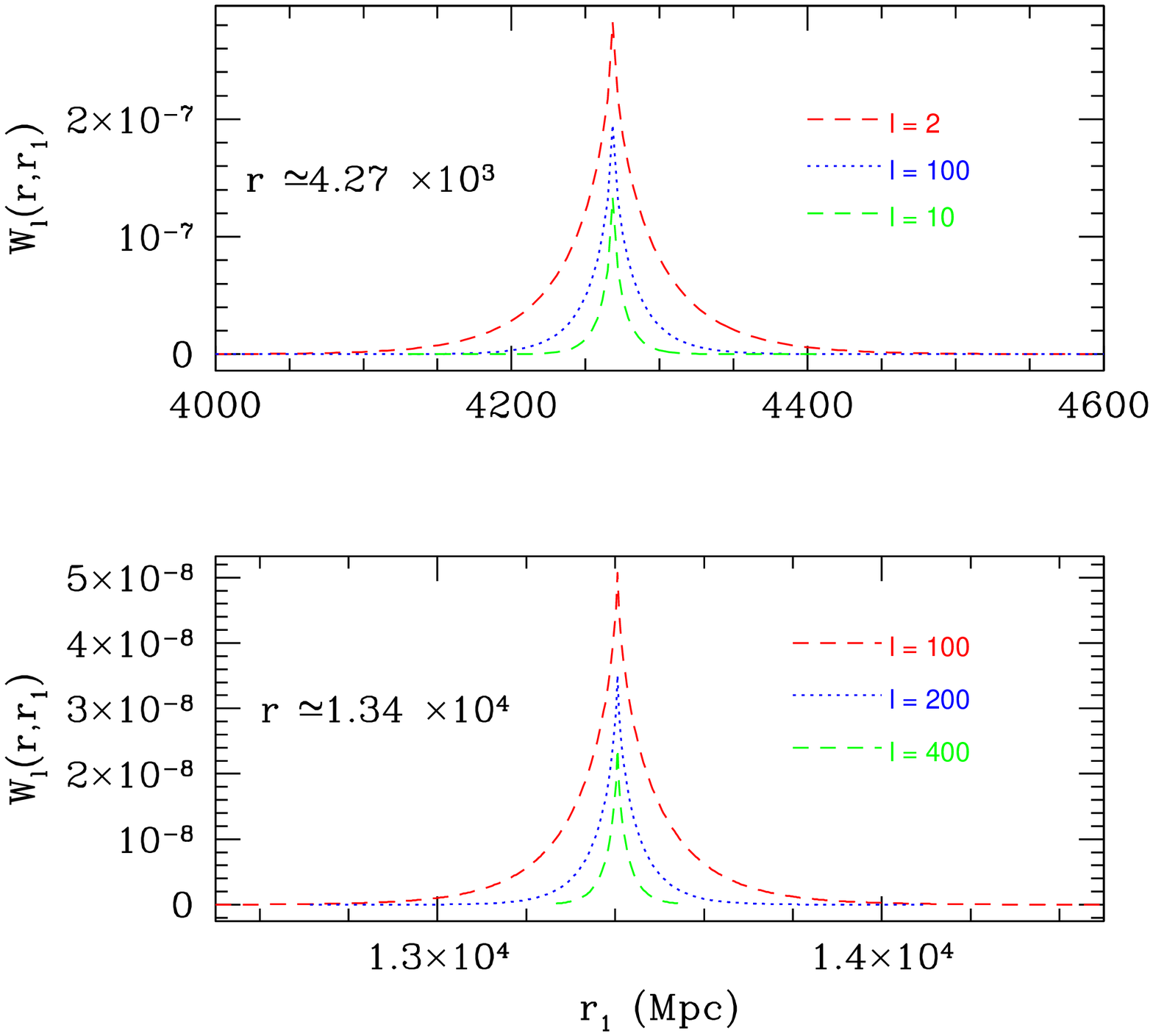}
\caption{Filter functions $W_\ell(r,r_1)$ plotted as a function 
of $r_1$ for two different fixed
  values of $r$. As $\ell$ gets larger, the $W_\ell(r,r_1)$ becomes
  more and more narrowly peaked around $r$.}\label{fig:wl2}
\end{center}
\end{figure}

In figures \ref{fig:wl1} and \ref{fig:wl2} we plot some $W_\ell$
functions for different values of $\ell$, $r$, $r_1$. As expected, 
$W_\ell(r,r_1)$ approximates a Dirac delta function
centered on $r$ with increasing accuracy for larger and larger
$\ell$. 
So for $l \gtrsim 10$ 
the coordinate $r_1$ needs to be sampled in a narrow region centered
around $r$. On the other hand, for low values of $\ell$,
$W_\ell(r,r_1)$ is non-negligible over a broad
$r_1$ range.
 Thus a coarser $r_1$ sampling over a larger $r_1$ interval is
required in this case. To check the accuracy of the numerical
computation of $W_\ell(r,r_1)$
it is useful to compute the angular power spectrum of $\Phi^L_{\ell
  m}(r)$ on a given spherical shell. Starting from formula
(\ref{eqn:nlm2phil}), and using the correlation properties of the
coefficients $n_{\ell m}(r_1)$ described by eqn. (\ref{eqn:whitenoise}) one gets:

\bea
\left\langle \Phi_{\ell_1 m_1}^L(x)  \Phi_{\ell_2 m_2}^{L*}(y) \right\rangle & = & 
\frac{2}{\pi}\delta_{\ell_1 \ell_2} \delta_{m_1 m_2} \int dr_1 dr_2
r_1^2 r_2^2 \left[ \langle n_{\ell_1 m_1}(r_1) n_{\ell_2 m_2}^*(r_2) \rangle
\right. \times \nonumber \\
 & & \left. \times \, W_{\ell_1}(x,r_1) W_{\ell_2}(y,r_2) \right] \nonumber \\ 
& = & \frac{2}{\pi} \int dr_1 dr_2 r_1^2 r_2^2
\frac{\delta^{(D)}(r_1-r_2)}{r_1^2} W_{\ell_1}(x,r_1)
W_{\ell_2}(y,r_2) \nonumber \\
& = & \frac{2}{\pi} \delta_{\ell_1 \ell_2} \delta_{m_1 m_2} \int dr_1
r_1^2 W_{\ell_1}(x,r_1) W_{\ell_2}(y,r_2) \; ,
\eea

which immediately yields:

\be\label{eqn:shellnormWl}
\langle |\Phi_{\ell m}^L(r)|^2 \rangle = \frac{2}{\pi} \int dr_1 r_1^2
W_\ell^2(r,r_1) \; .
\ee
  
Alternatively it is possible to use the following formula for the $\Phi_{\ell
  m}$ correlation function \cite{liguori}:
    
\be 
\left\langle \Phi_{\ell_1 m_1}^L(x)  \Phi_{\ell_2 m_2}^{L*}(y)
\right\rangle = \frac{2}{\pi} \delta_{\ell_1}^{\ell_2}
\delta_{m_1}^{m_2} \int dk k^2 P_\Phi(k) j_{\ell_1}(kx) j_{\ell_2}(ky)
\; ,
\ee

to find:

\be
\langle |\Phi_{\ell m}^L(r)|^2 \rangle = \frac{2}{\pi} \int dk k^2
 P(k) j^2_{\ell}(kr) \; .
\ee

For a primordial curvature perturbation power spectrum described by a power law
expression, $P(k) = Ak^{n-4}$, and using a well-known formula for the
Sachs-Wolfe effect, one finally gets:  

\be\label{eqn:shellnormgamma}
\langle |\Phi_{\ell m}^L(r)|^2 \rangle = 
\frac{2^{n-3} A \, r^{1-n}}{\pi}
     \frac{\Gamma \left(\ell+\frac{n}{2}-\frac{1}{2} \right) \Gamma
       \left(3-n \right)}{\Gamma \left(\ell+\frac{5}{2}-\frac{n}{2} \right)
       \Gamma^2 \left(2-\frac{n}{2}\right)} \; .
\ee

\begin{figure}[t!]
\begin{center}
\includegraphics[height=0.5\textheight, width = 0.7\textwidth]{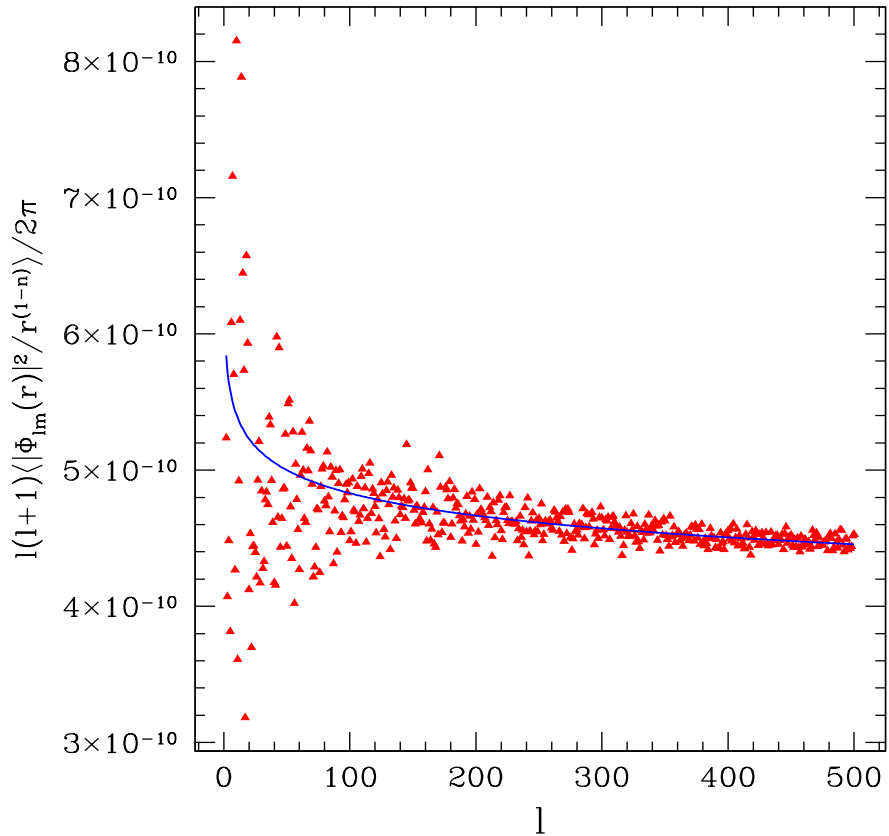}
\caption{Angular power spectrum of the Gaussian curvature perturbation multipoles
  $\Phi_{\ell m}^{\rm L}(r)$ obtained by averaging over all the
  spherical shells of a given simulation. In this example we consider a spectral index
  $n = 0.95$ and divide $|\Phi_{\ell m}^{\rm L}(r)|$ by $\sqrt{r^{(1-n)}}$
  in order to make the normalization of the spectrum
  independent of the shell radius before averaging. We compare the
  results extracted from our simulations (red triangles) to the expected shell power
  spectrum obtained from formula (\ref{eqn:shellnormgamma}), (blue line)}\label{fig:shellnorm}
\end{center}
\end{figure}

For a scale invariant primordial power spectrum one obtains, as expected,
$|\langle |\Phi_{\ell m}^L(r)|^2 \rangle| \propto {1/l(l+1)}$.
As we were anticipating above, one can use formulae (\ref{eqn:shellnormWl}) and
(\ref{eqn:shellnormgamma}) to test the $\Phi_{\ell m}(r)$ power
spectrum on different shells and the normalization of
$W_\ell(r,r_1)$. Results from our simulations are shown in picture
\ref{fig:shellnorm}. 

\begin{table}
\begin{center}
\begin{tabular}{|c|c|c|c|c|}
\hline
{\bf Noise} & {\bf Sky-cut} & $\mathbf{\langle f_{\rm NL}
  \rangle}$ & $\mathbf{\sigma_{maps}}$ & $\mathbf{\sigma_{fisher}}$ \\
\hline
No & No & $102.5$ & $11.1$ & $6.9$ \\
\hline
Homogeneous & No & $104.5$  & $15.8$ & $11$ \\
\hline
Homogeneous & $f_{sky} = 80$ & $105.2$ & $25.7$ & $12.2$ \\
\hline
\end{tabular}\caption{Results obtained from the application of the
  fast temperature $+$ polarization cubic statistics of \cite{Yadav} 
to a set of 300 non-Gaussian maps with an input $\fnl$ of $100$. First
column describes the noise properties of the map, second column is the
adopted sky-cut, third column is the average $f_{\rm NL}$ measured by the
estimators, fourth column is the measured $\fnl$ standard deviation,
fifth column is the expected standard deviation from a Fisher matrix 
analysis (i.e. neglecting corrections from the non-Gaussian part of
the multipoles).}
\label{tab:KSWresults}
\end{center}
\end{table}

\section{Fast cubic statistics and non-Gaussian maps}

\begin{figure*}
\begin{center}
\subfigure{\includegraphics[height=0.23\textheight,width =
    0.45\textwidth, angle=0]{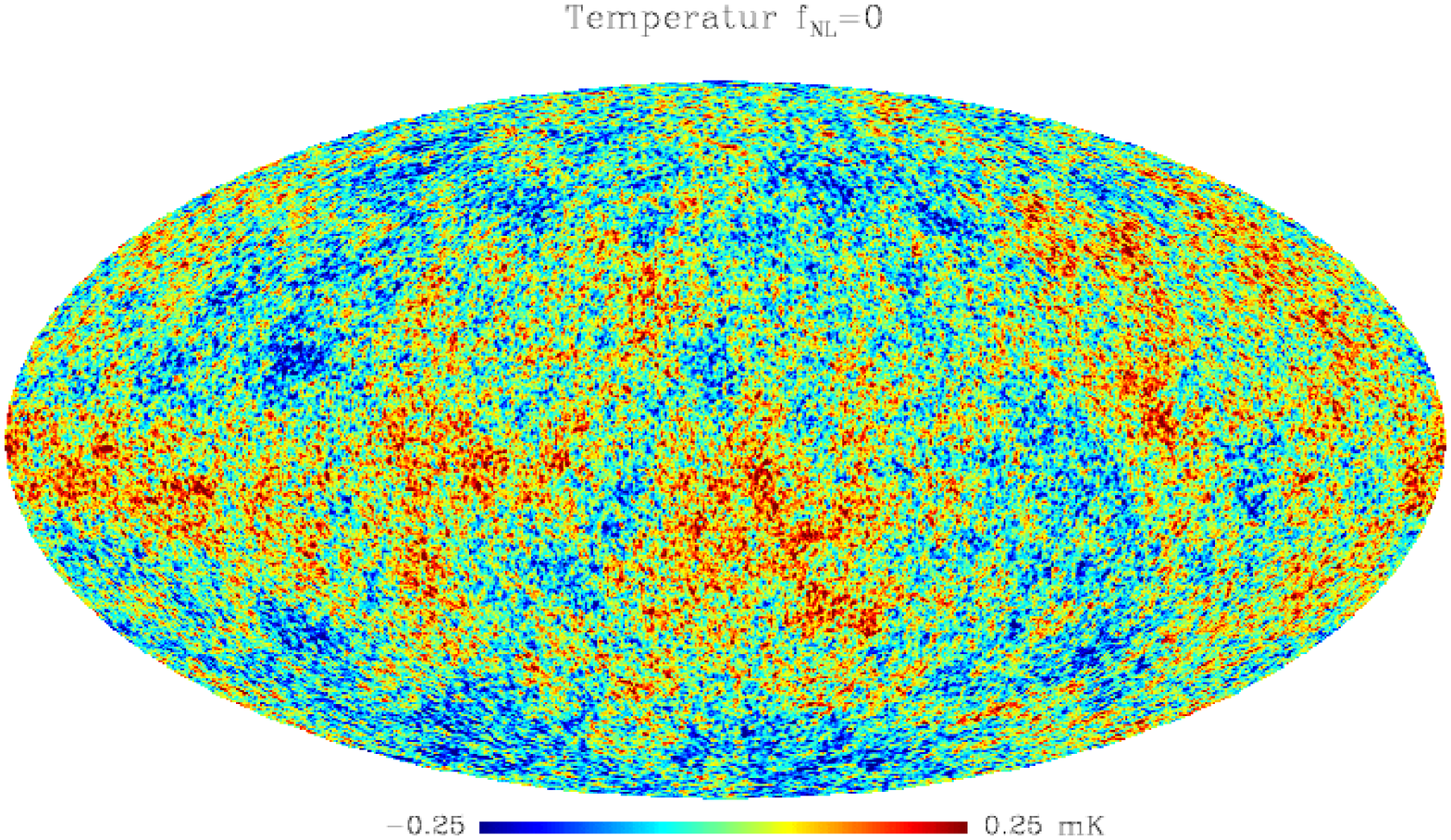}} \qquad
\subfigure{\includegraphics[height=0.35\textheight,width =
    0.3\textwidth,angle=90]{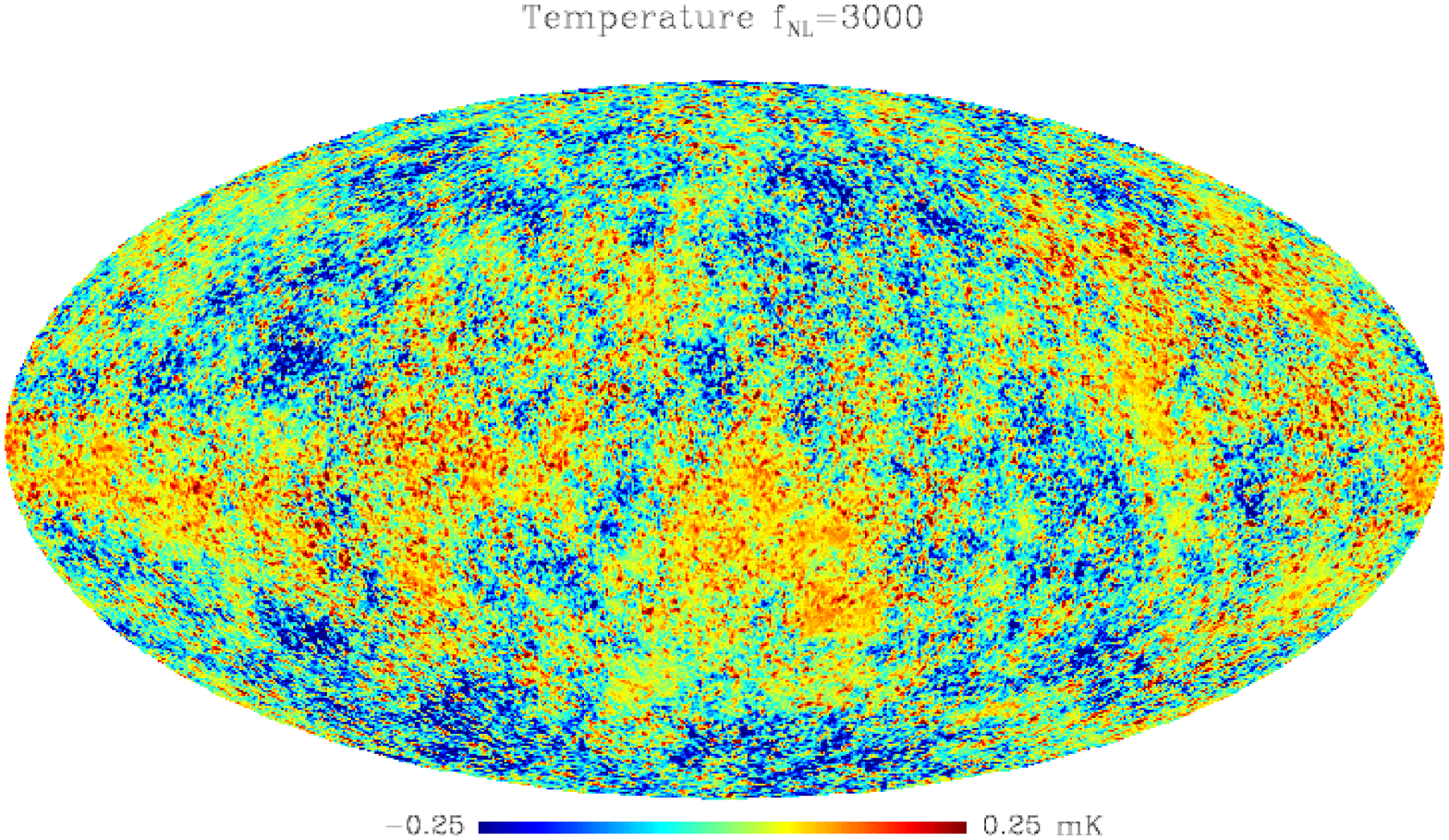}} \\ 
\subfigure{\includegraphics[height=0.35\textheight,width =
    0.3\textwidth,angle=90]{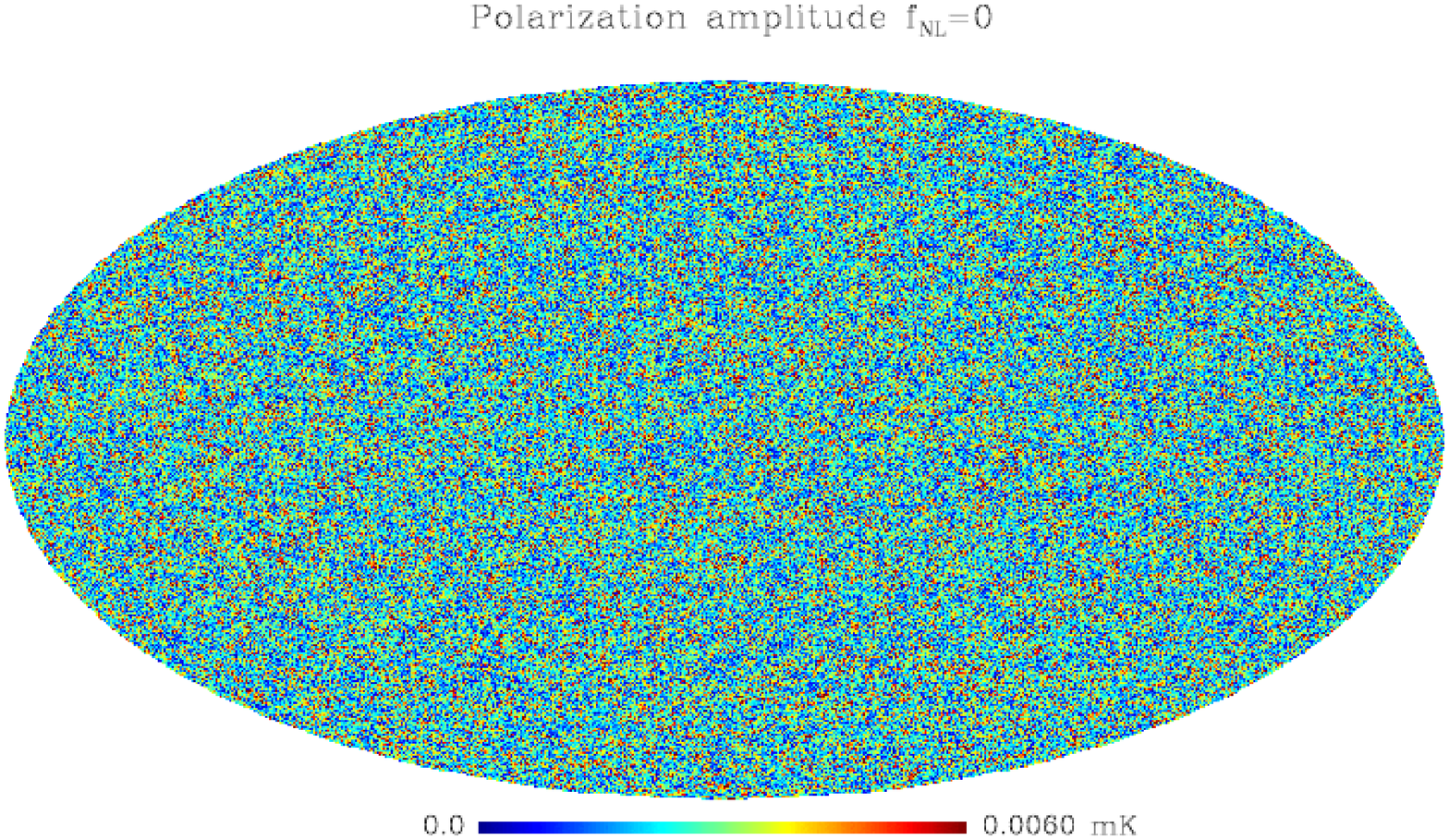}} \qquad
\subfigure{\includegraphics[height=0.35\textheight,width =
    0.3\textwidth,angle=90]{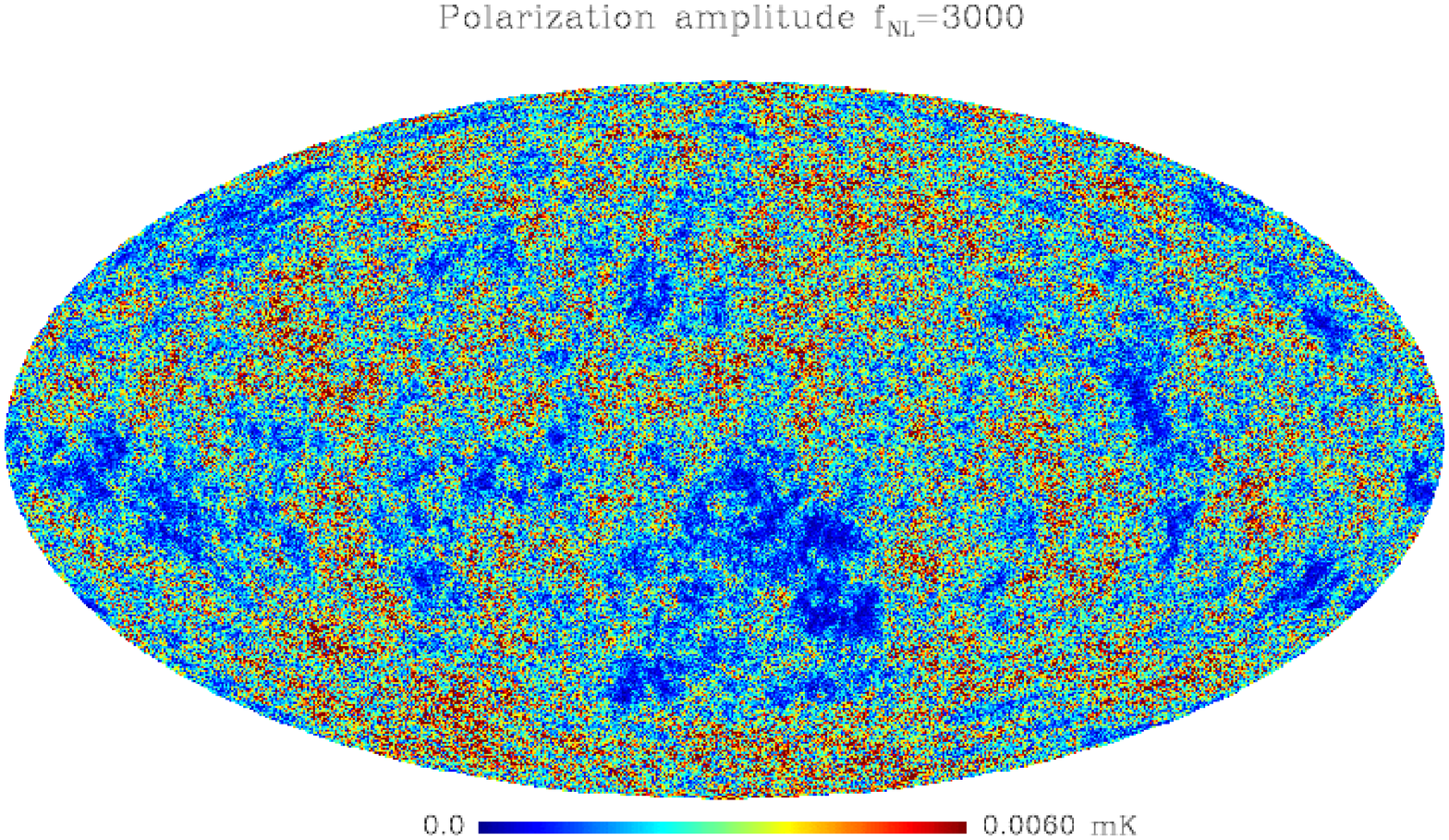}} 
\end{center}

\caption{Left column: temperature and polarization intensity Gaussian
  CMB simulations obtained from our algorithm. Polarization intensity
  is defined as $I \equiv \sqrt{Q^2 + U^2}$ where $Q$ and $U$
  are the Stokes parameters. Right column: temperature and polarization
  non-Gaussian maps with the same Gaussian seed as in the left column
  and $\fnl = 3000$. The reason for the choice of such a large $\fnl$
  is that we wanted to make non-Gaussian effects visible by eye in the
  figures. The cosmological model adopted for this plots is
  characterized by:  $\Omega_b = 0.042$, $\Omega_{cdm} = 0.239$, $\Omega_L
  = 0.719$, $h = 0.73$, $n = 1$, $\tau = 0.09$. Temperatures are in
  $mK$.}\label{fig:maps}
\end{figure*}

In order to test our algorithm we applied the temperature $+$
polarization fast-cubic statistics
described in \cite{Yadav} to a set of $300$
non-Gaussian simulations obtained from the cosmological
parameters $\Omega_b = 0.042$, $\Omega_{cdm} = 0.239$, $\Omega_L = 0.719$, 
$h = 0.73$
$n = 1$, $\tau = 0.09$. In figure \ref{fig:maps} we show a  
temperature and a
polarization intensity map extracted from this set.

When skycut is included a non-trivial correlation between large and
small $\ell$ is introduced. This correlation in turn produces a leakage of
power from high to low multipoles which tends to bias the estimator.
This effect has been accurately studied in \cite{Yadav}, where it has
also been shown that removing the lowest multipoles from the analysis
allows to circumvent this problem without a significant loss of
signal. For this reason the
first $30$ multipoles were not used in our analysis when a skycut was
considered. The exact $\ell_{min}$ was 
determined by preliminary applying the estimator to a set of Gaussian
simulation and estimating its variance as a function of $\ell_{min}$. We considered different 
sky cut levels and accounted for the presence of homogeneous noise. Our
results are summarized in table \ref{tab:KSWresults} .

Our computation provides evidence for the unbiasedness of the estimator
but shows at
the same time a discrepancy between the calculated error bars and
Fisher matrix based expectations (note that these expectations are
obtained at zeroth order in $\fnl$, thus neglecting $\fnl$-dependent
terms in the three point function). These discrepancies are $\fnl$
dependent: for small undetectable $\fnl$ we find a good agreement
between Fisher matrix estimates and our results whereas increasing
values of  $\fnl$ produce larger and larger differences. This effect
had been predicted and explained by the authors of
\cite{Creminelli2}. It arises from $\fnl$ dependent correction terms
in the variance of the estimator. These terms become important when 
$\fnl$ is detected at several sigma. The comparison of  our
results with those in \cite{Creminelli2} is necessarily approximate 
because the latter were
obtained in the flat-sky approximation and ignoring radiation
transfer functions. However we can still cross-check for a qualitative 
agreement between the two results. Using the above approximations, 
the $\fnl$-dependent formula describing the estimator variance is:

\be
\sigma^2  = \langle \sigma^2 \rangle_{\fnl = 0} \left(
1 + \frac{8 \fnl^2 A N_{pix}}{\pi \ln N_{pix}} \right) \; ,
\ee

where $ \langle \sigma^2 \rangle_{\fnl = 0}$ is the estimator variance
in the Gaussian case (i.e. the variance estimated from the Fisher
matrix), $A$ is the amplitude of primordial perturbations and
$N_{pix}$ is the number of pixels in the map. To simplify the notation
we define $\sigma^2_0 \equiv \langle \sigma^2 \rangle_{\fnl = 0}$. 
Following \cite{Creminelli2} we consider an $\fnl$ detection at
$n\sigma_0$. From the formula above:

\be
 \sigma^2  = \sigma_0^2 + \frac{2 n^2
  \sigma_0^2}{\pi \ln^2 N_{pix}} \; .
\ee

We then find the expected relative correction to the variance as:

\be\label{eqn:relativecorr}
\frac{\langle \sigma^2 \rangle}{\sigma_0^2} -1 = \frac{2 n^2}{\pi \ln^2 N_{pix}} \; .
\ee

\begin{figure}[t!]
\begin{center}
\includegraphics[height=0.4\textheight, width = 0.8\textwidth]{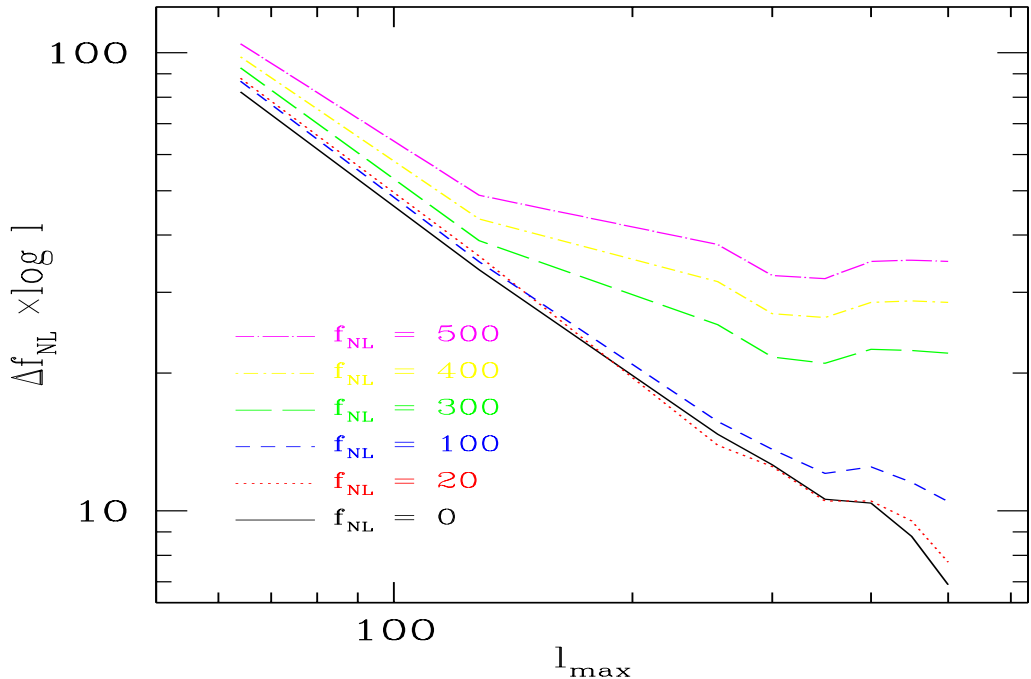}
\caption{Error bars estimated from different sets of simulations
  including various $\ell_{lmax}$ and input $\fnl$. The error bars are
  compared to the corresponding Fisher matrix forecast. As explained in the text, 
  an $\fnl$-dependent correction to the estimator variance make 
  the error bars to scale as ${1/\ln \ell_{max}}$ instead of
  ${1/\ell_{max}}$ when $\fnl$ is large enough to produce a several 
  sigma detection at a given angular resolution.}\label{fig:sigmascaling}
\end{center}
\end{figure}

In our analysis we have $N_{pix} = 3145728$ (HEALPix $nside =512$) and
$\sigma_0 = 6.9$ for the case without  sky-cut or noise (see second
line of table \ref{tab:KSWresults}). We have an input $\fnl$ of $100$, so
this corresponds to $n = 14.5$. Plugging this numbers into the equation
above we obtain a relative correction of $0.6$ which is about one
third of the
observed ${\sigma^2/\sigma_0^2 -1 } = 1.6$ but in  qualitative
agreement considering the approximations contained in
eqn.~(\ref{eqn:relativecorr}). For large enough $\fnl$
eqn.~(\ref{eqn:relativecorr}) also
predicts the variance of the estimator to decrease as ${1/\ln^2 N_{pix}
\sim {1/\ln \ell_{max}}}$, much slower than the Fisher matrix forecast of 
$\sigma \sim {1/\ell_{max}}$. We explicitly tested this prediction on sets
of simulated maps with different $\fnl,N_{pix},\ell_{max}$ and we found
a good agreement between theory and simulations, as depicted in figure
\ref{fig:sigmascaling}. Thus the results
obtained analytically in \cite{Creminelli2} under several simplified 
assumptions are confirmed by our numerical approach, which works in
full-sky and includes radiation transfer functions.     
 
\section{Computational requirements and possible applications} 

Our algorithm takes about $3$ hours on a normal PC 
to generate a map with $\ell_{max} = 500$, 
$N_{pix} \simeq 10^6$, corresponding to an analysis at 
WMAP angular resolution.
The most time consuming part is the computation of the harmonic
transforms required to generate $\Phi^{\rm NL}_{\ell m}$ from  
$\Phi^{\rm L}_{\ell m}$. As we generate the primordial curvature perturbation in
about $400$ spherical shells we need to make $400$ calls to the 
HEALpix synfast
and anafast subroutines respectively. Thus we can roughly quantify the 
CPU time for a non-Gaussian simulation at a given resolution as the
time required to produce $800$ Gaussian maps at the same resolution. 
It is thus clear that the generation of maps at the resolution
achieved by {\em Planck} constitutes a very intensive computational
task and requires a parallelization of the algorithm. Only the
temperature version of the code has been parallelized so far, enabling
us to generate a map at $\ell_{\max} = 3000$, $nside = 2048$ in about
$2$ hours on $60$ processors. A set of $300$ temperature 
maps with this angular resolution has been generated and
tested. Extending the parallel code in order to include polarization
should be straightforward, because all the sampling-related problems have
been already solved for the serial version of the algorithm presented
in this paper and including polarization transfer functions is
trivial. 
The total CPU time to generate a map is going to be
unchanged with respect to the temperature-only version, because the
primordial curvature perturbation generation scheme is  identical and the total 
number of shells is basically the same. We would like to note here
that a different algorithm has been proposed for the
generation of non-Gaussian maps in \cite{Smith}. This algorithm can
generate maps with a given two and three point function but does not
reproduce the higher order correlation functions predicted by the model. By
making this approximation, the authors of \cite{Smith} are able to
dramatically speed up the computation ($\sim 3$ minutes for a map at 
$\ell_{max} = 1000$ on a single processor). In the limit of weak
non-Gaussianity cite{note1}  neglecting higher
order correlation functions should be a good approximation. In
particular 
it has been explicitly shown in \cite{Creminelli2} that no
additional information on $\fnl$ can be added by applying estimators
based on higher order correlators. This conclusion is strictly related
to the presence of $\fnl$-dependent correction terms in the variance
of the local bispectrum.
Note however that these terms have originally been studied in
flat-sky approximation and neglecting transfer functions. As an
application of our algorithm, in the previous section of this paper 
we have explicitly cross-checked the results 
of \cite{Creminelli2} using our simulations which are full-sky and
account for radiative transfer \cite{note2}

We would also like to stress that being 
able to correctly reproduce higher order correlation functions in the
simulations was fundamental in order to make this test. The reason is
that what we are studying here is actually an $\fnl$-dependent
correction to the 6-point function (bispectrum variance) coming from 
a product of the 2-point function with the 4-point function (see again
\cite{Creminelli2} for further details). 

Another obvious application for these simulated maps is given by the
possibility 
to use them in order to test and calibrate not only the bispectrum but 
any kind of estimator (like e.g. Minkowski functionals, wavelets and
so on). 
In particular the analytical formulae of the Minkowski functionals
recently derived by \cite{hikage} may be compared with our
simulations of the temperature maps. 
Our preliminary investigation shows a very good agreement,
which gives us further confidence in the accuracy of the simulated
temperature maps.
Despite the optimality of the bispectrum just discussed above, using
different estimators is still important, especially in view of a possible $\fnl$ 
detection by Planck. Alternative estimators should in fact be used in
this case in order to cross-validate such detection.   
 
Furthermore, it is interesting to notice that the algorithm we are describing is 
not only able to generate non-Gaussian CMB maps, but it also produces
maps of the primordial curvature perturbation $\Phi(r,\hat{r})$,
sampled in the
relevant radii for the generation of the final CMB signal. This allows us
to apply and test tomographic reconstruction techniques of the
curvature perturbation like those proposed in \cite{tomography}. This
will be the object of a forthcoming publication \cite{yadavinprep}. 
Finally we would like
to observe that the same elegant $r$-sampling optimization technique
introduced in \cite{Smith} can be implemented in our case in order 
to drastically 
reduce the number of radii in which the primordial curvature perturbation must be
evaluated. Following the results of \cite{Smith},
a good accuracy in
the final maps should be obtained using only $20$ spherical shells in
our code after this optimization. As we are now using $400$ shells, we
estimate a speed improvement of a factor $\sim 20$. In this way the
parallel version of the algorithm should allow the generation of a map
at full Planck resolution in $\sim$ $10$ minutes against the present
$2$ hours. For this reason CPU time does not seem to be a problem and
tests of non-Gaussianity at {\em Planck} angular resolution using
our algorithm are perfectly feasible.

\section{Conclusions}  

In this paper the algorithm for the generation of
non-Gaussian primordial CMB maps originally introduced in
\cite{liguori} has been
generalized by including a polarization component in the
simulations. 
Using this generalized algorithm 
we have produced a set of $300$ temperature and polarization
maps at WMAP angular resolution. We have then
analyzed these simulations using the fast cubic temperature $+$ polarization
statistics recently introduced by the authors of \cite{Yadav}. We have
verified that we can extract the correct input $\fnl$ from the maps,
thus checking at the same time both the unbiasedness of the estimator and 
the reliability of the simulations. We also studied the estimator
variance on different sets of maps including various angular resolutions and
input $\fnl$. We found that an $\fnl$-dependent
correction to the estimator variance induces a discrepancy between
the error bars extracted from the simulations and the Fisher matrix
estimate of the same error bars at $\fnl = 0$. We therefore confirmed 
previous findings by the authors \cite{Creminelli2}. At the same
time, differently from previous approaches, our 
numerical Monte Carlo analysis allowed us to work in full sky and
account for radiation transfer functions. We finally discussed future 
applications of our
simulations, which will include a detailed analysis of non-Gaussian temperature
and polarization simulations at Planck angular resolution.

\acknowledgments{We acknowledge the use of the HEALpix software
  \cite{gorski1, gorski2} (see http://healpix.jpl.nasa.gov/ for
  further information on HEALpix). We acknowledge partial financial
  support from from the ASI contract Planck LFI Activity of Phase E2. 
  We would like to thank Paolo Cabella for
  stimulating discussions and contributions in an early phase of this
  project. We would also like to thank Paolo Creminelli for useful
  discussions. ML is supported by PPARC.}


\begin{thebibliography}{}

\bibitem{acqua}
V.~Acquaviva, N.~Bartolo, S.~Matarrese and A.~Riotto,
Nucl.\ Phys.\ B 667 (2003) 119, [arXiv:astro-ph/0209156] 

\bibitem{maldacena}
J.~Maldacena, JHEP\ 0305 (2003) 013, [arXiv:astro-ph/0210603] 

\bibitem{Lyth} D.H. Lyth, C. Ungarelli, D. Wands, Phys.Rev. D67 (2003)
  023503, [arXiv:astro-ph/0208055] 

\bibitem{BMR} N. Bartolo, S. Matarrese, A. Riotto, 
JHEP 0404 (2004) 006, [arXiv:astro-ph/0308088]

\bibitem{komrev} N. Bartolo, E. Komatsu, S. Matarrese, A. Riotto,
Phys.Rept. 402 (2004) 103-266, [arXiv:astro-ph/0406398]
 
\bibitem{Ginfl} N. Arkani-Hamed, P. Creminelli,
  S. Mukohyama, M. Zaldarriaga, JCAP 0404 (2004) 001, [arXiv:hep-ph/0312100]

\bibitem{DBI1} M. Alishahiha, E. Silverstein, D. Tong Phys.Rev. D70 (2004) 123505,  
[arXiv:hep-th/0404084]

\bibitem{DBI2} X. Chen, Phys.Rev. D72 (2005) 123518, [arXiv:astro-ph/0507053]

\bibitem{Shellard} G.I. Rigopoulos, E.P.S. Shellard, B.J.W. van Tent, [arXiv:astro-ph/0511041]

\bibitem{multifield} L.E. Allen, S. Gupta, D. Wands, 
JCAP 0601 (2006) 006, [arXiv:astro-ph/0509719]
 
\bibitem{Sefusatti} E. Sefusatti and E. Komatsu, [arXiv:0705.0343]

\bibitem{Pillepich} A. Pillepich, C. Porciani, S. Matarrese,
  Astrophys.J. 662, 1 (2007) 1-14, [arXiv:astro-ph/0611126] 

\bibitem{Cooray} A. Cooray, Phys.Rev.Lett. 97 (2006) 261301,
  [arXiv:astro-ph/0610257]

\bibitem{NGWMAP1} E. Komatsu, et al., Astrophys.J.Suppl. 143 (2003) 119

\bibitem{NGWMAP3} D.N. Spergel, et al., Astrophys.J.Suppl. 170 (2007)
	377

\bibitem{Creminelli2} P. Creminelli, L. Senatore, M. Zaldarriaga, JCAP
  0703 (2007) 005, [arXiv:astro-ph/0606001]

\bibitem{KS2001} E. Komatsu and D. Spergel, Phys.Rev. D63 (2001)
  063002, [arXiv:astro-ph/0005036] 

\bibitem{BabichZalda} D. Babich, M. Zaldarriaga, Phys.Rev. D70 (2004)
  083005, [arXiv:astro-ph/0408455]

\bibitem{Yadav} A. P. S. Yadav, E. Komatsu, B. D. Wandelt,
  arXiv:astro-ph/0701921

\bibitem{KSW} E. Komatsu, B. Wandelt, D. Spergel, Astrophys.J. 634
  (2005) 14-19, [arXiv:astro-ph/0305189]

\bibitem{CreminelliKSW}  P. Creminelli, A. Nicolis, L.
  Senatore, M. Tegmark, M. Zaldarriaga, JCAP 0605 (2006) 004

\bibitem{Cornish} N.J. Cornish, D.N. Spergel, G.D. Starkman, E. Komatsu,
Phys.Rev.Lett. 92 (2004) 201302
  
\bibitem{liguori} M. Liguori, S. Matarrese, L. Moscardini, 
Astrophys.J. 597 (2003) 57-65, [arXiv:astro-ph/0306248] 

\bibitem{Smith} K. M. Smith, M. Zaldarriaga, [arXiv:astro-ph/0612571]

\bibitem{hikage} C. Hikage, E. Komatsu, T. Matsubara,
Astrophys. J., 653 (2006) 11

\bibitem{tomography} A. P. S. Yadav, B. D. Wandelt, 
Phys.Rev. D71 (2005) 123004, [arXiv:astro-ph/0505386] 


 
\bibitem{yadavinprep} A. P. S. Yadav {\em et al.}, in preparation

\bibitem{gorski1} K.M. Gorski, et al., Astrophys.J. 622 (2005)
  759-771, [arXiv:astro-ph/0409513 ]

\bibitem{gorski2} K.M. Gorski, et al., [arXiv:astro-ph/9905275]

\bibitem{note1} This limit is verified in our case as non-Gaussianity from inflation is small.

\bibitem{note2} As a subject of future work, all these checks will be repeated by taking into account
possible contaminant effects, like e.g. foreground residuals, 
second order anisotropies, systematics,
map-making effects and so on, in order to study their impact on the estimator.

\end{thebibliography}
\end{document}